# Controlling Communication Field of Complex Networks by Transformation Method

Xiangyang Lu[1,2], Jin Hu[1,a], Ran Tao[1,b], and Yue Wang[1]

1. School of Information and Electronics, Beijing Institute of Technology,
Beijing 100081, People's Republic of China
2. School of Electronic and Information Engineering, Zhongyuan University of Technology
Zhengzhou 450007, People's Republic of China

Corresponding authors: [a] bithj@bit.edu.cn, [b] rantao@bit.edu.cn

**Abstract**: Controlling the global statuses of a network by its local dynamic parameters is an important issue, and it is difficult to obtain the direct solution for. The transformation method, which is originally used to control physical field by designing material parameters, is proposed to obtain the necessary local dynamic parameters when the global statuses of a network system are prescribed in a space. The feasibility of this transformation method is demonstrated and verified by two examples (a communication field cloak and a communication field bender) in the network system. It is shown that the global system state can be controlled by adjusting the local nodes dynamics with the transformation method. Simulation results also show that the transformation method provides a powerful, intuitive and direct way for the global statuses controlling of network systems.

1. Introduction

Research on complex networks is prevalent in recent years [1-19]. The various evolution models of complex networks were mainly developed from two aspects, i.e., the dynamics *of* networks and the dynamics *on* network [1]. The dynamics *of* networks explores the network topology dynamics change over time. Its typical models include regular graph model, random graph model [7], small world model [8,9,10] and scale free networks [11,12]etc. The dynamics *on* network describes state characteristic of network nodes, such as the synchronization of networks [13,14], communication [15], controlling of dynamic networks [16-19], etc.

Many dynamics on network focuses on the system state change in temporal domain. A typical case is the epidemic spreading in a network system [8], which figured out the probability behavior of the nodes being infected in a period. This result has also been verified in the epidemic spreading in small world [9]. Recently researches on the complex network dynamic in temporal-spatial domain received many attentions. Paper [20] proposed using heat conduction equation to describe recommendation mechanism on a social network. The similar mechanism also was introduced in [21]. In [22, 23], it was proposed that the information communication field spreads throughout the system via a diffusion-like process.

Once the dynamic mechanism of the network is established, it is another interesting issue to control the network behaviors. In [16, 17] the pinning control of the networks is presented as a demonstration. A spatio-temporal chaos control is further developed to enhance the relatively weak chaos through pinning control

[19,24]. In [25] it is also shown that the small world network's dynamics can still be maintained through key nodes protection.

Another important network control is to design the local network spatial statuses, so that the entire system status may have desired global patterns or configurations, related to some performances or functions. It is a typical inverse problem to design the global statuses by controlling or adjusting the local dynamic characteristics. Generally, a direct solution is difficult to be obtained. In this paper, the transformation method [26-28] is proposed to resolve such problems, provided that the dynamic equation on network is form-invariant under coordinate transformations. The method has been successfully applied in physical field control. Moreover we will show that the method is a powerful, intuitive and direct way for the network controlling by taking the controlling of communication field in a network as an example.

We start in section 2 by reviewing the governing equation of communication field and the importance of communication field control. In section 3 the transformation method is introduced and proposed to control the communication field. Thereafter the necessary local dynamic parameters are obtained to achieve the controlling, followed by two examples in section 4. In section 5 we give discussions and conclusions.

2. Communication field in a network

In order to consider the spatial effects on a physical space of a social network, the scalar spatiotemporal communication field is introduced in [22]. The opinions can spread throughout the system via a diffusion-like process and will influence the individual's action. A similar idea has also been explored in [20], where the heat conduction mechanism is proposed as a recommended mechanism in a social network system. In general form, the governing equation of communication fields $u(\mathbf{x},t)$ in a two-dimensional spatial system can be written as

$$\rho \frac{\partial}{\partial t} u(\mathbf{x},t) = \nabla \alpha \nabla u(\mathbf{x},t) - \beta u(\mathbf{x},t) + f(\mathbf{x}), \tag{1}$$

where $\mathbf{x}$ presents spatial coordinates, $t$ is time, $\rho$ is the density, $\alpha$ is the diffusion coefficient, $\beta$ is the production or absorption coefficient and $f(\mathbf{x})$ denotes the sources. In [22], $u(\mathbf{x},t)$ represents one of the communication fields $h_\theta(\mathbf{x},t)$ ($\theta = \pm 1$), the density is normalized to be 1, and the diffusion coefficient $\alpha = D_h$ determines the speed of any information received by the individual; the diffusion coefficient is a constant thus there is $\nabla \alpha \nabla u(\mathbf{x},t) = D_h \Delta u(\mathbf{x},t)$; the decay rate $\beta$ determines the duration of the existence of the generated information and the sources term is $f(\mathbf{x}) = \sum_{i=1}^{N} s_i \delta_{\sigma,\sigma_i} \delta(\mathbf{x}-\mathbf{x}_i)$, which describes the individuals contribution to the filed. In [20], Eq. (1) is simply reduced to Laplace's equation, which describes the steady state of the heat conduction.

Once the governing equation of the communication field is established, the spatio-temporal distributions can be obtained. It is achieved by applying appropriate boundary conditions and initial conditions. Also distribution of the communication field is very important due to its impact on the following behavior of the network. For example, in the collective opinion system [22], depending on the local information, the individual may either change its opinion or migrate towards a location which provides a larger support to its current opinion. Especially, in artificial network system, such as wireless sensor networks, municipal transportation network, multi-robots systems (MRS) etc, either in transient state or steady state, the distribution of communication field is very important to control the global configuration and behavior of the system. Since the communication field is the solution of the governing equation (1), in order to distribute the communication field as prescribed global pattern, we have to redesign the parameters ($\rho$, $\alpha$, $\beta$ and $f$) in every point to fulfill the requirement.

However, it is a typical inverse problem to obtain the global distribution of the communication field by adjusting the local dynamic parameters of the network system, which usually involves complex mathematical technique due to non-unique solution. Recently in physics community, the so-called transformation method [26-28] has been proposed to circumvent this difficulty. It can directly construct one solution for the above inverse problem. For the sake of completeness, we first briefly review the main idea of transformation method, and then apply it to communication field controlling.

3. Controlling communication field by transformation method

3.1 Transformation method

The transformation method is based on the form-invariance of the corresponding governing equation under coordinate transformations [26-28]. Consider a specific dynamic phenomenon in a region $\Omega$, which is governed by a system of differential equations $F$ under a Cartesian frame,

$$F[\mathbf{x}, t, \mathbf{C}(\mathbf{x}), \mathbf{u}(\mathbf{x}, t)] = 0, \quad \mathbf{x} \in \Omega, \tag{2}$$

where $\mathbf{C}$ and $\mathbf{u}$ represent all the involved dynamic parameters and the dependent fields, respectively. Equation (2) describes the specific dynamic mechanisms at any point **x** in $\Omega$ region and characterizes the intrinsic relation between the fields and parameters.

If the space is spanned with any curved coordinates defined by $\mathbf{x}' = \mathbf{x}'(\mathbf{x})$, the form-invariance of equation (2) indicates that the governing equations will retain their form in the new coordinate system of new region $\Omega'$, i.e., equation operator $F$ still works

$$F[\mathbf{x}', t, \mathbf{C}'(\mathbf{x}'), \mathbf{u}'(\mathbf{x}', t)] = 0, \quad \mathbf{x}' \in \Omega', \tag{3}$$

where $\mathbf{C}'$ and $\mathbf{u}'$ are the transformed parameters and fields interpreted in the new coordinate system

$$\mathbf{C}'(\mathbf{x}') = T_C[\mathbf{C}(\mathbf{x})], \quad \mathbf{u}'(\mathbf{x}', t) = T_u[\mathbf{u}(\mathbf{x}, t)], \tag{4}$$

and completely determined by the spatial mapping transformation $\mathbf{x}' = \mathbf{x}'(\mathbf{x})$. The form-invariant transformation indicates that a field $\mathbf{u}(\mathbf{x}, t)$ distributed on the original space $\Omega$ will be mapped to $\mathbf{u}'(\mathbf{x}', t)$ on another space $\Omega'$ with the same time variable $t$, and the dynamic mechanisms are still being kept. Therefore, one can control the spatial distribution of the fields in an intuitive and direct way by the mapping $\mathbf{x}' = \mathbf{x}'(\mathbf{x})$ such that the transformed field $\mathbf{u}'$ in the new space follows a prescribed way, and the first term of Eq. (4) can give necessary parameters for this function. Before demonstrating this method with two examples, we will show that Eq. (1) is form-invariant and obtain the expressions in Eq. (4).

3.2 Transformation of communication field equation

Rewrite Eq. (1) as

$$\nabla \alpha \nabla u(\mathbf{x},t) = g(\mathbf{x},t), \tag{5}$$

where

$$g(\mathbf{x},t) = \rho \frac{\partial}{\partial t} u(\mathbf{x},t) + \beta u(\mathbf{x},t) - f(\mathbf{x}), \tag{6}$$

Equation (5) has the same form as conductivity equation or Helmholtz's equation and known as the form-invariant equation [29], it means that, after the mapping transformation $\mathbf{x}' = \mathbf{x}'(\mathbf{x})$ from space $\Omega$ to space $\Omega'$, equation (5) can retain its form with respect to the new coordinates $\mathbf{x}'$ in the new space $\Omega'$, that is

$$\nabla' \boldsymbol{\alpha}' \nabla' u'(\mathbf{x}',t) = g'(\mathbf{x}',t), \tag{7}$$

and the transformations of the parameters and filed are [29]

$$\boldsymbol{\alpha}' = \frac{\mathbf{A}\alpha\mathbf{A}^{\mathrm{T}}}{\det \mathbf{A}}, \quad g' = \frac{g}{\det \mathbf{A}}, \quad u' = u, \tag{8a}$$

where $\mathbf{A}$ is the Jacobian transformation tensor with the components $A_{ij} = \partial x'_i / \partial x_j$. From Eq. (6) and (8a), we can further obtain the other transformations of the parameters

$$\rho' = \frac{\rho}{\det \mathbf{A}}, \quad \beta' = \frac{\beta}{\det \mathbf{A}}, \quad f' = \frac{f}{\det \mathbf{A}}. \tag{8b}$$

Now both the communication field and the parameters can be transformed from one space to another, which is still controlled by Eq. (1). Then, we can design specific space mapping to characterize specific function by redistributing the communication field. The necessary parameters for the new fields distribution can be obtained by Eq. (8).

4. Examples of communication field controlling by transformation method

In order to control the communication field by transformation method, one should associate the desired function with a fictitious space which has some topological feature, and is mapped from a flat Cartesian space. If the space mapping is given, then the Jacobian transformation tensor **A** can be obtained, in turn of the necessary parameters from Eq. (8). To illustrate this method, we will give two examples, one is the communication field cloak and the other is communication field bender.

4.1 Communication field cloak

Suppose there is a special area $\Sigma$ in the system and the communication field is required can not diffuse into it, meanwhile the diffusion of communication field in the other area is required can not changed by this special strategy. In that case, the special area is in some sense "invisible" to other area. The simplest way for the non-diffusion of $\Sigma$ is to use an "insulation blanket" to surround it; however, this blanket will influence the diffusion of communication field in rest of the area according to the governing equation (1), thus the invisibility of $\Sigma$ is lost. In order to keep $\Sigma$ invisible, we will design a so-called "invisible cloak", which can guide communication field around $\Sigma$, and return them to their original trajectory, as a result the outer area of cloak can not be disturbed by this rounding [27]. In polar coordinate system, the linear radial transformation of the cloak is

$$r' = a + (b-a)r/b, \quad \theta' = \theta, \qquad (9)$$

where $a$ and $b$ are the inner and outer radii of the cloak, respectively, and area $\Sigma$ is inside the cloaked area where $r' < a$, as shown in Fig. 1. According to Eq. (8), the necessary parameters for this cloak are

$$\alpha'_r = \frac{r'-a}{r'}\alpha, \quad \alpha'_\theta = \frac{r'}{r'-a}\alpha,$$
$$\rho' = (\frac{b}{b-a})^2 \frac{r'-a}{r'}\rho, \qquad (10)$$

$$\beta' = (\frac{b}{b-a})^2 \frac{r'-a}{r'} \beta,$$
$$f' = (\frac{b}{b-a})^2 \frac{r'-a}{r'} f.$$

In the cloak body $b \geq r' \geq a$, all the parameters become inhomogeneous and the diffusion coefficient becomes anisotropic. Parameters in other area are unchanged.

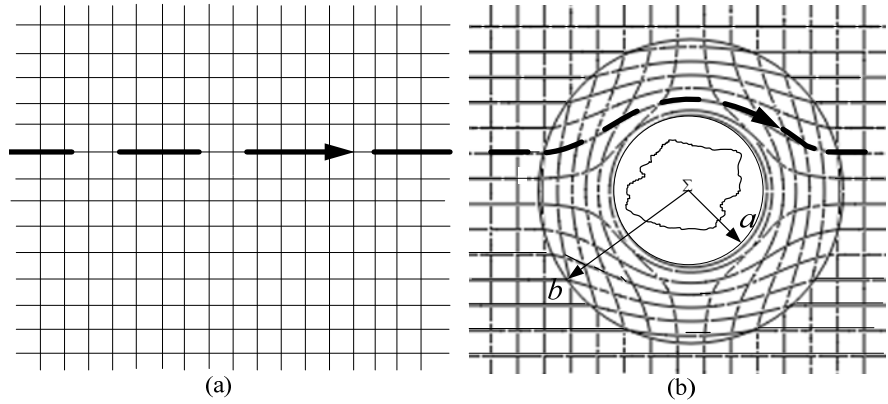

Figure 1. The space transformation of the communication field cloak.

(a) The original space $\Omega$, (b) the transformed space $\Omega'$.

Figure 2 shows the computational domain in which the area I is a subsystem where the nodes have sources and area II is a subsystem where the nodes have no sources, and therefore the communication field will diffuse from area I to area II. The area $\Sigma$ located in area II needs to be cloaked.

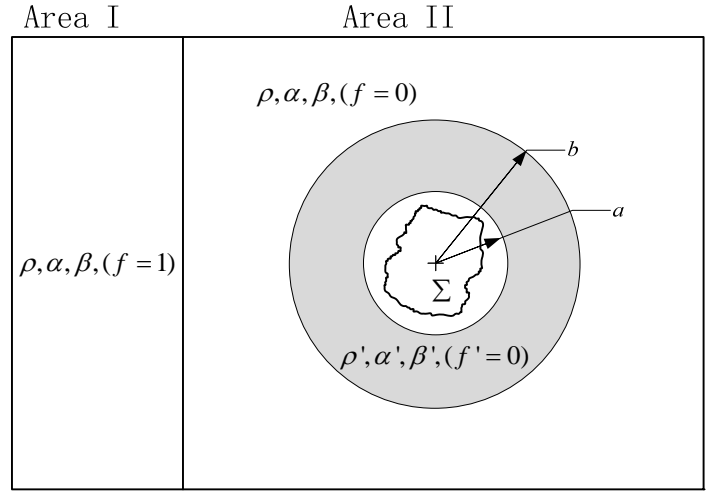

Figure 2.　The computational domain of communication field cloak.

Figure 3 shows the simulation results of the communication field cloak in using COMSOL Multiphysics finite element-based PDE solver, where all original non-zero parameters are normalized to be 1. It is shown that, without the designed cloak but with the insulation blanket, the whole outside distribution of communication field will be changed as compared to the original system (Fig. 1b). However, with the help of the designed cloak, the cloaked area is invisible, i.e., the outside communication field is the same as that is in the original system (Fig. 3 c and d).

It is worth pointing out that the perfect cloak requires some components of the parameters to be infinite at the inner boundary $r'=a$, see $\alpha'_\theta$ in Eq. (10). This singularity can be avoided by the technique proposed in [30]. In addition, the cloak can be in arbitrary shape, as shown in [31].

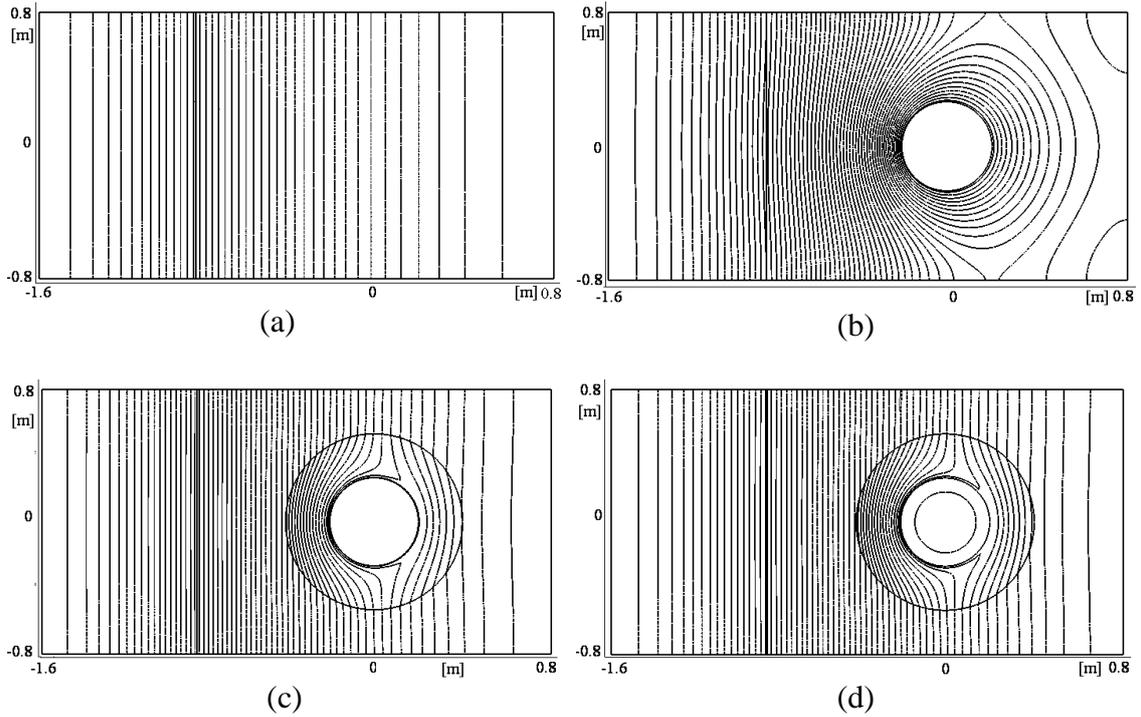

Figure 3. The simulation results of communication field cloak. The lines in the computational domain are the contour plot of the communication field $u$ in (a) the original system, (b) the system with only insulation blanket surrounding the cloaked area, (c) the system with designed cloak at time $t=1$, and (d) the system with designed cloak in steady state.

4.2 Communication field bender

The communication field bender is used for the synchronization of the communication field diffusion in curve path. Usually, the diffusion ray of communication field in an irregular area depends on the shape of the boundaries. Thus it is difficult to control the synchronization of the communication field diffusion. For example, in an arc shape system as shown in Fig. 4, an input communication field $u_{\text{input}}$ exists at the boundary AD; and for the homogeneous parameters, the communication field contour is curve lines, which means that the communication field asynchronously diffuses to boundary CB. In order to make the communication field synchronously arrive at boundary CB, we use transformation method to design a communication field bender, it is constructed by a spatial transformation that makes a rectangular plate with side lengths $ka$ and side width $a$ respectively transformed into a arc shaped area with a polar angle $\phi$, as shown in Fig. 5.

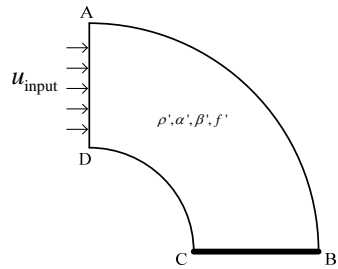

Figure 4. The computational domain of communication field bender.

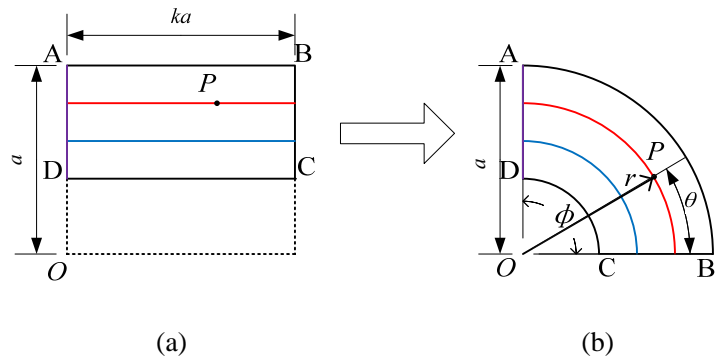

(a)             (b)

Figure 5. Transformation of a communication field bender.

(a) The original space $\Omega$, (b) the transformed space $\Omega'$.

From Eq. (8), the designed system has the anisotropic diffusion coefficient. However, by the technique proposed in [32] for isotropic bender which sets the stretch along $\hat{\mathbf{r}}$ equals to that along $\hat{\boldsymbol{\theta}}$ at each point, the transformed parameters then become

isotropic

$$\alpha' = \alpha,$$
$$\rho' = \frac{r\pi}{2a}\rho, \quad (11)$$
$$\beta' = \frac{r\pi}{2a}\beta,$$
$$f' = \frac{r\pi}{2a}f,$$

when $\phi = \pi/2$ and $k=1$. According to transformation method, the original straight communication field contour will still be kept straight. The simulation verifies this design, as shown in Fig. 6, where the system is in steady state. The input communication field diffuses in the bender with straight contour, and it arrives at boundary BC at the same time.

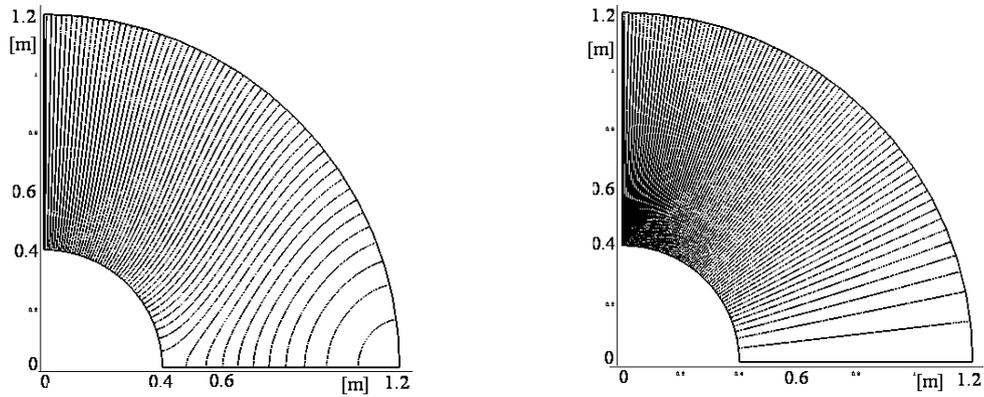

Figure 6. The simulation results of communication field bender.

The lines in the computational domain are the contour plot of the communication field $u$ in (a) the original system with parameters are normalized to be 1 and (b) the designed communication field bender with parameters of Eq. (11).

5. Discussions and conclusions

The designing and controlling of the global statuses of a network system is an important issue for some artificial network system. It can be further used to implement the living self-organization mechanism of an information system, and will make the system more flexible when deal with the environmental modifications [34]. Although the network is usually physically implemented as discrete system, the concept of field and the related continuous physical methods are often used in network theory with some approximations [1,7,23]. In this point of view, it is necessary and important to extend the transformation method to network control.

In summary, we propose the transformation method to control the global statuses of network. The transformation method provides a powerful, intuitive and direct way for the network controlling, if the dynamics of the system expressed by a set of partial differential equation with form-invariant feature. The method is demonstrated and verified by two examples of communication field control in network systems, one is communication field cloak and the other is a communication field bender, both of which can have potential practical applications. The method only controls the spatial statuses of the network. However, compared with physical materials, the artificial systems are more convenient to design time varying local parameters. Therefore the transformation method may also be very important in designing and controlling temporal statuses of the networks, which currently is under investigation.

**Acknowledgment:** This work was supported by the National Natural Science Foundation of China ( 11172037).